# How oil slicks floating on the ocean affect SST?

**Kejing. Liu**[1]

[1] Navigation Institute, Jimei University; Xiamen, China.

**Key Points:**

- An analytical dynamic system model is proposed to explain how oil slicks make potential increase on sea surface temperature
- Oil slicks make the uncertainty increase and lower the predictability of models on climate change.

**Abstract**

Oil slicks are widely distributed in the ocean today,as a kind of coverage on sea surface , they became a part of ocean environment and affect their surroundings. A stochastic-dynamic theoretical model proposed in this work to illustrate how oil slicks affect global climate from micro scale relation between a piece of oil slick and sea surface temperature (SST) of its surrounding unit area, for SST is an important index of global climate. The model indicate that oil slicks make the sea surface warmer in the future, and the temperature series of the sea surface covered by oil slicks will have greater variance and fatter tails for its distribution and reduce SST predictability from a microcosmic perspective. Thus, more oil infused into the ocean makes the air–sea system more uncertain. These findings indicate that the present air–sea coupled models may lack of sufficient attention to oil slicks floating on the sea surface.

**1 Introduction**

The air–sea coupled models focused on large-scale oscillators have revealed the conjugate relationship between the ocean and atmosphere, energy transition between interface of them regulating global climate, sea surface temperature (SST) is an important indicator of ocean energy debuget. Observations and models have shown that SST has increased over the past few decades and seems to have accelerated in approximately 50 years (USEPA, 2024). Many lage scale factors, e.g., circulations, have been known for influence SST change. Hasselmann (1976) divides the climate system into two subsystems, a rapidly varying system and a slowly responding system; the latter produces a cumulative response to the former and builds a stochastic model that reveals the mechanisms of climate change. Circulations, is a long term change part of the climat system composed with the ocean and atmosphere, however, for complete description of cumulative influence of sustainably and varying processes on the whole system, we need attention to the phenomenons with smaller scale and with faster change speed. Satellites have depicted a clear picture of the global ocean in which almost all the coastlines of the world, especially in nearshore areas or the shipping lanes with heavy traffic, are scattered with oil slicks, the area is larger than previous estimate and most of the slicks are from anthropogenic discharges (Dong, Liu, Hu, et al. 2022). With industrate developing, effects of the human actives, e.g., shipping, on climate and ocean ecosystem becoming unneglible, ship emissions into atmosphere make cloud brightening and affect climate through make more aerosols and increase the number of clould droplets (M. S. Diamond, H. M. Director, R. Eastman, et al. 2020). As another type of petrochemical emissions, oil released into water synchronously with aerosols into atmosphere, both of them belongs to rapidly varying factors of climate system. As coverge on the sea surface, oil slicks make the interface between atmosphere and ocean different with the natural sea surface, they may affect energy transition. Illustrate how oil slicks affect the atmosphere-ocean interface is useful to understand the influence of anthropogenic discharges on global change in small spatial and temporal scales.

**2 Surface temperature of oil slicks and SST**

From bulk parameterization (J. A. Businger, 1975), the SST variation equation (K. Hasselmann, 1976; Li, 1981) can be written as

$$h\rho^S C_p^S \frac{\partial T_S}{\partial t} = H_S^S + H_L^S + H_R^S \tag{1}$$

where $h$ is the thickness of thermocline, $\rho^S$ is the sea water density, $C_p^S$ is the specific heat at constant pressure of sea water, $T_S$ is the SST, $H_S^S$ is the sensible heat flux (SHF), $H_L^S$ is the latent heat flux (LHF), $H_R^S$ is the radiation flux (RF).

From it can obtain

$$\frac{\partial T_S}{\partial t} = \frac{1}{h} \cdot \frac{1}{\rho^S C_p^S} \cdot \left( H_S^S + H_L^S + H_R^S \right) \tag{2}$$

The heat fluxes on the right side are divided into long-term averages, represented by $\langle \cdot \rangle$, and random anomalies, represented by variables with an apostrophe,

$$\frac{\partial T_S}{\partial t} = \frac{1}{h} \cdot \frac{1}{\rho^S C_p^S} \cdot \left[ \langle H_S^S + H_L^S + H_R^S \rangle + H_S'^S + H_L'^S + H_R'^S \right] \tag{3}$$

it can be written as

$$T_S(t + \delta t) = T_S(t) + \frac{1}{h} \cdot \frac{1}{\rho^S C_p^S} \cdot \left[ \langle H \rangle + H' \right] \tag{4}$$

where $\delta t$ is the time step or a determined interval.

Under an area covered with a piece of oil slick, the slick is in direct contact with the atmosphere before its complete decomposition, not the water surface, although it is very thin (Fig. 1). Thus, its surface temperature can be written as

$$T_S^O(t + \delta t) = T_S^O(t) + \frac{1}{h^O} \cdot \frac{1}{\rho^O C_p^O} \cdot \left[ \langle H^O \rangle + H'^O \right] \tag{5}$$

where $T_S^O$ is the surface temperature of the area covered with oil slick, $h^O$ is the thickness of the slick, $\rho^O$ is the oil density, $C_p^O$ is the specific heat at constant pressure of oil, $H^O$ represents the total heat flux between the slick and atmosphere.

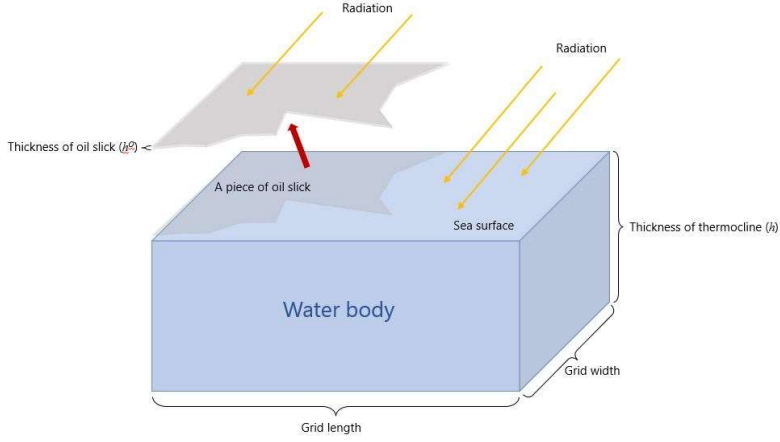

**Fig. 1.** Conceptual map of a unit (one grid) of sea water covered with a piece of oil slick. The blue part represents the unit of the water body; its area is an arbitrarily given length plus an arbitrarily given width, and its thickness is the thickness of the thermocline locally. The gray part represents the oil slick floating on the sea surface, which is very thin. The yellow arrows represent radiation from atmosphere, for the part covered with the oil slick, they do not contact directly with the sea surface but with the slick.

Assuming that, at the present time, on the unit area, temperature of the part covered with oil slick is equal to that of the natural sea surface, $T_S(t) = T_S^O(t)$. From the bulk formula (J. A. Businger, 1975) of heat flux

$$H_L + H_S = C_H(1+B)\rho C_p (T - T_S) \cdot |V| \quad (6)$$

(where $C_H$ is the transfer coefficient, $B$ is the Bowen ratio, $T$ is air temperature, $T_S$ is surface temperature (sea or slick), $|V|$ is the absolute value of the difference between wind speeds of a reference height in atmospheric surface layer and at the surface, on very small scale it is the maximum surface component of turbulence) and the formula of black body radiation

$$H_R = \frac{1}{4}\mu I_0 (1-\alpha) - \varepsilon \sigma T_S^4 \quad (7)$$

(where $\mu$ is the extrinsic parameter, $I_0$ is the solar constant, $\alpha$ is the albedo of the surface, $\varepsilon$ is the radiance of the surface, $\sigma$ is the Stefan-Boltzman constant), the second term of the right of equation (4) or equation (5) can be written as

$$\frac{1}{h}\cdot\frac{1}{\rho C_p}\cdot\left[\langle H\rangle + H'\right] = \frac{1}{h}C_H(1+B)(T-T_S)\cdot|V| + \frac{1}{h}\cdot\frac{1}{\rho C_p}\left[\frac{1}{4}\mu I_0(1-\alpha) - \varepsilon\sigma T_S^4\right] + \frac{1}{h}\cdot\frac{1}{\rho C_p}H'$$

(8)

It is obviously that $h^O \ll h$, $\rho^O < \rho^S$, $C_p^O < C_p^S$, thus

$$\frac{1}{h^O}C_H(1+B)(T-T_S(t))\cdot|V| \gg \frac{1}{h}C_H(1+B)(T-T_S(t))\cdot|V| \qquad (9)$$

For $\varepsilon^S > \varepsilon^O$, $\alpha^O > \alpha^S$, thus

$$\frac{1}{4}\mu I_0(1-\alpha^S) > \frac{1}{4}\mu I_0(1-\alpha^O) \qquad (10),$$

and $\varepsilon^S \sigma T_S^4 > \varepsilon^O \sigma T_S^4 \qquad (11),$

the two sides of equation (10) and equation (11) are in same order of magnitude but $\frac{h\rho^S C_p^S}{h^O \rho^O C_p^O} \propto 10^6$ for unit of $h$ is m and of $h^O$ is μm, so

$$\frac{1}{h^O}\cdot\frac{1}{\rho^O C_p^O}\left[\frac{1}{4}\mu I_0(1-\alpha^O)-\varepsilon^O \sigma T_S^4\right] \gg \frac{1}{h}\cdot\frac{1}{\rho^S C_p^S}\left[\frac{1}{4}\mu I_0(1-\alpha^S)-\varepsilon^S \sigma T_S^4\right] \qquad (12)$$

From equation (9) and equation (12), it is clearly

$$\frac{1}{h^O}\cdot\frac{1}{\rho^O C_p^O}\langle H^O\rangle \gg \frac{1}{h}\cdot\frac{1}{\rho^S C_p^S}\langle H\rangle \qquad (13)$$

For the random anomalies $H'$ and $H'^O$, they present rapidly varying parts of the system, consider them random signals, using the Fourier transform relation and obtain

$$H'(t) = \frac{1}{2\pi}\int_{-\infty}^{\infty}\hat{H}(\omega)\cdot e^{i\omega t}d\omega \qquad (14)$$

where $\hat{H}(\omega)$ is the Fourier transform coefficient and $\omega$ is the frequency of the signal.

The power spectral density (PSD) of the signal

$$P(\omega) = |\hat{H}(\omega)|^2 \propto \omega^{-\beta} \qquad (15)$$

where $\beta$ is the power spectral index, for Brownian motion, $\beta = 2$; for Lévy flight, $\beta < 1$

The variance of $H'$ is

$$Var[H'] = \int_{-\infty}^{\infty}P(\omega)d\omega \qquad (16)$$

Thus, if $\omega_1 > \omega_2$, $P(\omega_1) < P(\omega_2)$, and $Var[H_1'] < Var[H_2']$ exist.

Notice that signal of $H'^O$ (denoted as $\hat{H}(\omega^{OS})$) can be decomposed into natural micro processes, such as turbulence of sea surface ($\hat{H}(\omega^S)$) and the frequency of oil discharge ($\hat{H}(\omega^O)$), as follows:

$$\hat{H}(\omega^S) = A^S \cos(\omega^S t + \varphi^S) \qquad (17)$$

$$\hat{H}(\omega^O) = A^O \cos(\omega^O t + \varphi^O) \qquad (18)$$

where $A$ is the amplitude of the signal, $A^S \neq A^O$; $\omega^S \neq \omega^O$; $\varphi$ is the phase, $\varphi^S \neq \varphi^O$. Their composition is as follows:

$$\begin{aligned}\hat{H}(\omega^{OS}) &= A^S \cos(\omega^S t + \varphi^S) + A^O \cos(\omega^O t + \varphi^O) \\ &= (A^S + A^O) \cos\left[\frac{|\omega^S - \omega^O|}{2} t + \frac{|\varphi^S - \varphi^O|}{2}\right] \cdot \cos\left[\frac{\omega^S + \omega^O}{2} t + \frac{\varphi^S + \varphi^O}{2}\right]\end{aligned} \qquad (19)$$

where the combined amplitude is $A^{OS} = (A^S + A^O) \cos\left[\frac{|\omega^S - \omega^O|}{2} t + \frac{|\varphi^S - \varphi^O|}{2}\right]$, which changes regularly, and the frequency of its change is $\omega^{OS} = \frac{|\omega^S - \omega^O|}{2\pi}$, it is obvious that $\omega^{OS} < \omega^S$. If the values of $\omega^S$ and $\omega^O$ are similar, periodic changes of $A^{OS}$ can be called beats, its frequency $\omega^{OS}$ will be very small.

Thus, regardless of whether the signals of $H'$ and $H'^O$ are Brownian motions or Lévy flights (there is a hypothesis: the signal of $H'^O$ is more likely to be close to Lévy flight than that of $H'$), there must have

$$Var[H'^O] > Var[H'] \qquad (20)$$

From it, can obtain

$$P[H'^O > H'] > P[H' > H'^O] \qquad (21)$$

If the signal of $H'^O$ is Lévy flight, whereas that of $H'$ is Brownian motion, this inference will be stronger.

This means distribution function of $H'^O$ has a fatter tail than that of $H'$, and is more likely to be longer and have a lager extreme value.

From equation (4), equation (5), equation (13) and equation (21), when $T_S(t) = T_S^O(t)$, with a high probability,

$$T_S^O(t+\delta t) > T_S(t+\delta t) \quad (22)$$

This is how oil slicks on the sea surface increases SST at a very small scale. On a unit area of sea surface (can be $0.1° \times 0.1°$), part of it covered with a piece of oil slick, denote the whole SST is $T_{ST}$, it can be written as

$$\frac{\partial T_{ST}}{\partial t} = \left(1 - \frac{S^O}{S}\right)\frac{\partial T_S}{\partial t} + \frac{S^O}{S} \cdot \frac{\partial T_S^O}{\partial t} \quad (23)$$

where $S$ is the total area and $S^O$ is the area of oil slicks in the grid, so $\frac{S^O}{S}$ is normalized area of oil slicks per grid. From equation (22), it is easy to obtain

$$T_S^O(t+\delta t) > T_{ST}(t+\delta t) > T_S(t+\delta t) \quad (24),$$

If $\frac{S^O}{S} \to 1$, $T_{ST} \to T_S^O$ exists, which means that with more oil slicks released into ocean, the SST tend to be higher.

## 3 System predictability

Hasselmann (1976) proposed a method for climate system predictability:

$$p(y, t) = (2\pi)^{-\frac{n}{2}} |R|^{-\frac{1}{2}} \times \exp\left(-\frac{R_{ij}^{-1}}{2}(y_i - \langle y_i \rangle)(y_j - \langle y_j \rangle)\right) \quad (25)$$

which is a general solution of the Fokker-Planck equation of the general stochastic climate model, where $y$ represents the long-term response part of the whole system, $p(y, t)$ represents the probability density distribution of climate states in the climatic phase space, and $R_{ij}$ represents the covariance matrix of $y$. The evolution of $p(y, t)$ determines the degree of climate predictability, and the predictive skill of model can be defined as

$$s = \min(s_1, s_2) \quad (26),$$

$$s_1 = \frac{\delta_1}{\left(r^2 + \delta_1^2\right)^{\frac{1}{2}}} \quad (27),$$

$$s_2 = \frac{\delta_2}{(r^2 + \delta_2^2)^{\frac{1}{2}}} \quad (28),$$

where $r$ is the RMSD of every state and predicted climate state (mean state), $\delta_1$ is the distance from the mean climate state and initial state, $\delta_2$ is the distance from the mean climate state and asymptotic state, which can be regarded as a very long-term stationary mean state. Both of them must be defined in terms of some suitable positive definite matrix $M_{ij}$ so that $s$ is a simple number.

For $p(T_S, t)$ and $p(T_S^O, t)$, let the present time be the initial time, and $T_S(t) = T_S^O(t)$ can be rewritten as $T_S(0) = T_S^O(0)$, in other words, the initial states of $T_S$ and $T_S^O$ are the same. In addition, in the present AOGCMs or air-sea coupling models, the sea surface is taken as a uniform material surface, so their predicted states (mean states) are the same, $\langle T_S \rangle = \langle T_S^O \rangle$, and asymptotic states (long term mean states) are also the same, $\langle T_S \rangle_\infty = \langle T_S^O \rangle_\infty$, so

$$\delta_1^O = \delta_1^S \quad (29),$$

$$\delta_2^O = \delta_2^S \quad (30),$$

From equation (22),

$$r^O > r^S \quad (31),$$

then form equation (27), equation (28) and equation (26),

$$s_1^O < s_1^S \quad (32),$$

$$s_2^O < s_2^S \quad (33),$$

$$s^O < s^S \quad (34).$$

That means that the sea surface is covered by oil slicks with lower predictability in present models (Fig. 2, modified from the Fig.3 of Hasselmann, K. (1976), with unrestricted use permit), in another word, oil slicks make models predictability dacay faster.

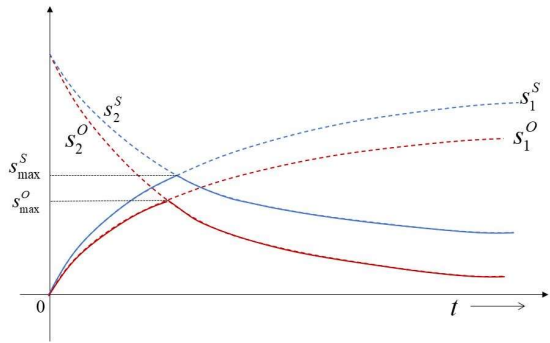

**Fig. 2.** Schematic of system predictability. The blue lines represent the predictive skill indices $s_1^S$ and $s_2^S$ of the model to the SST, the solid parts compose the index $s^S$, and its peak represents its maximum, which is also the maximum predictability that the model can reach for the SST. The red lines represent the predictive skill indices $s_1^O$ and $s_2^O$ of the model for the temperature of the sea surface covered with oil slicks, the solid parts constitute the index $s^O$, and its peak represents the maximum value, which is also the maximum predictability of the model can reach for the oil slick temperature. The blue solid line is above the red solid line, which means that the predictability of the model to the surface of the oil slick will never reach that of the model to the SST after the initial time and will continue to decline after its peak value; even though the line representing $s^S$ has the same trend, it remains higher than $s^O$.

## 4 Conclusions

To understand how continuous and varying processes effect the long term part of climate system helps the comprehension on the total geosystem. Oil slicks are scattered coverage on the sea surface, especially in nearshore areas and shipping lanes with high density traffic, most of them are origin from anthropogenic discharges. On the sea surface area covered with them, the thin petrochemical slick separate atmosphere and natural sea surface.

This work forcus on physical and maethmatical model of the effect of every slick on a very small area of the sea surface and the spatial and temporal cumulative effects on climate represented by substaintial change in the SST. As a kind of very thin coverage, oil slicks increasing the SST and variance in the SST, they decrease the predictability of the SST in the present models. With high emission and intensive transportation, more oil slicks on the sea surface, which makes the SST incontrollable and may present more unpredictable extreme values or inaccurate long-term predictions, more uncertainty makes more complex in the system. The predictability of a system with two (or more) media is not a simple sum of that of the media, the real sea surface dotted with oil slicks is a complex system, the importance of ocean oil slicks in air–sea systems should be given sufficient attention for global sustainable development.

For discuss random anomalies of the heat flux on the sea surface or the oil slick, this work take them as signals, and derived that oil slick increase variance of the random anomalies and make them increase in high probability. In the process of derivation, we propose a hypothesis that is the signal of random anomalies of surface with oil slick on it is more likely to be close to Lévy flight than that of natural sea surface, which haven't verified but do not affect the result of this work. However, stochastic features may affect degree of the effect of oil slicks on sea surface, they may also be related to types or properties of petrochemical oil, these are left for future research.

Furthermore, similar to oceans covered by oil slicks regionally, land surfaces covered by asphalt, concrete and other artifacts are grouped or scattered in the areas with human living. In areas where artifacts have high density, the artificial coverings may play similar roles in the ground–atmosphere system as oil slicks in the air–sea system, this may be a extending of the theory proposed in this work.

**Acknowledgments**

- The author thanks the United States Environmental Protection Agency (EPA) for providing data of average global sea surface temperature, 1880-2023 and data of change in sea surface temperature, 1901-2022, and thanks for the data of the authors of "Chronic oiling in global oceans" sharing in the supplementary materials of their paper.

- This research was supported by Jimei University grant for National Fund Cultivation Program ZP2023005